\documentclass{article}
\usepackage{spconf,amsmath,graphicx}
\usepackage{booktabs}
\usepackage{hyperref}

\usepackage{xcolor}
\newcommand{\qingming}[1]{\textcolor{green}{ #1 (Qingming)}}
\newcommand{\chiehchi}[1]{\textcolor{red}{ #1 (Chieh-Chi)}}
\newcommand{\viktor}[1]{\textcolor{blue}{ #1 (Viktor)}}
\renewcommand{\qingming}[1]{}
\renewcommand{\chiehchi}[1]{}
\renewcommand{\viktor}[1]{}
\usepackage[normalem]{ulem}
\newcommand{\floor}[1]{\left\lfloor #1 \right\rfloor}

\title{Weight-sharing Supernet for Searching Specialized Acoustic Event Classification Networks Across Device Constraints}
\name{Guan-Ting Lin$^{\dagger12}$\thanks{$^{\dagger}$This work was done during Guan-Ting’s internship at Amazon.}, Qingming Tang$^{2}$, Chieh-Chi Kao$^{2}$, Viktor Rozgic$^{2}$, Chao Wang$^{2}$}
\address{National Taiwan University$^{1}$      Amazon Alexa$^{2}$ }

\begin{document}
\ninept
\maketitle
\begin{abstract}
 Acoustic Event Classification (AEC) has been widely used in devices such as smart speakers and mobile phones for home safety or accessibility support \cite{lopatka2016detection}. As AEC models run on more and more devices with diverse computation resource constraints, it became increasingly expensive to develop models that are tuned to achieve optimal accuracy/computation trade-off for each given computation resource constraint. In this paper, we introduce a Once-For-All (OFA) Neural Architecture Search (NAS) framework for AEC. Specifically, we first train a weight-sharing supernet that supports different model architectures, followed by automatically searching for a model given specific computational resource constraints. Our experimental results showed that by just training once, the resulting model from NAS significantly outperforms both models trained individually from scratch and knowledge distillation (25.4\% and 7.3\% relative improvement). 
 We also found that the benefit of weight-sharing supernet training of ultra-small models comes not only from searching but from optimization.

\end{abstract}
\vspace{-0.3cm}
\section{Introduction}
Acoustic Event Classification (AEC) is the task of detecting the occurrence of certain events based on acoustic signals.
It has been widely used in devices such as smart speakers and mobile phones for home safety or accessibility support \cite{lopatka2016detection}.
Previous works have shown great progress on AEC using Convolutional Neural Network (CNN) \cite{pann, hershey2017cnn, xu2017convolutional, takahashi2016deep} and Audio Spectrogram Transformer (AST) \cite{ast, ssast, paast, maeast, audioMAE}. 
However, such models are usually computationally expensive and not suitable for edge devices (e.g., 86M \qingming{Number of parameters may vary due to implementation? I think we specifically mean the model from this paper \cite{ast}?}model parameters for AST~\cite{ast}).
To address this issue, several \textit{Model Compression} methods have been proposed using \textit{Knowledge Distillation} \cite{choi2022temporal, gao2022multi, jung2020knowledge, shi2019compression, wu2018reducing, shi2019teacher, gong2022cmkd, Shi2019}, which distills the learned knowledge from teacher model to a small student model.


As AEC models run on more and more devices with diverse computation resource constraints, one emerging area of research is how to efficiently train models with optimal accuracy/computation trade-off for each given computation resource constraint. For a given constraint, Neural Architecture Search (NAS) can be used to find the optimal model architecture. However, this still requires a dedicated training process for each given constraint, which is inefficient for developing models for devices with diverse computation resource constraints. 
One-shot weight-sharing NAS approaches \cite{ofa, compofa, alphanet, autoformer} were proposed to solve this problem by training a weight-sharing supernet that contains various sub-networks and then searching for the specialized sub-network given diverse resource constraints without re-training, which reduces the training effort for $O(N)$ devices from $O(N)$ to $O(1)$.
Although this approach had shown great performance on image recognition tasks~\cite{ofa, compofa, alphanet, autoformer}, there is no previous work using weight-sharing NAS in the audio domain. The most relevant work is LightHuBERT \cite{lighthubert}, which adopts the once-for-all \cite{ofa} framework on the large transformer-based model. However, it mainly focuses on self-supervised and task-agnostic speech representation learning.



This work is the first attempt to efficiently train models with optimal accuracy/computation trade-off for each given computation resource constraint on AEC. 
Specifically, we first leverage knowledge distillation to train a weight-sharing CNN-based AEC supernet (``Once-for-all" supernet) that contains sub-networks with shared weights. After supernet training, we directly search for specialized sub-network given a specific constraint without re-training. 
We choose CNN-based architectures for our experiments since it requires less computational resource than AST-based architectures, making them more suitable for edge devices. Experiments are conducted on AudioSet \cite{audioset}, which is a popular multi-labeled audio classification benchmark. Our contributions can be summarized below: 
\vspace{-0.1cm}
\begin{itemize}
    \item To the best of our knowledge, this is the first work to efficiently train models for edge devices with diverse computation resource constraints using weight-sharing NAS for AEC. Although this work focuses on AEC, the underlying design could potentially be applied to other audio applications like music genre classification and keyword spotting.
    \item Experimental results show that the proposed method can find specialized AEC sub-networks that significantly outperform models trained from scratch, also outperforming models trained with knowledge distillation (25.4\% and 7.3\% relative improvement).  
    \item We found that adopting elastic depth (Section \ref{sec:2-2})\qingming{Should we refer readers to section 2.2?}in the first few blocks is harmful to the AEC supernet. Using only elastic depth in the last few blocks results in a good trade-off between model performance and computation cost.  
\end{itemize}


\begin{figure*}[t]
  \centering
  \includegraphics[width=0.85\linewidth]{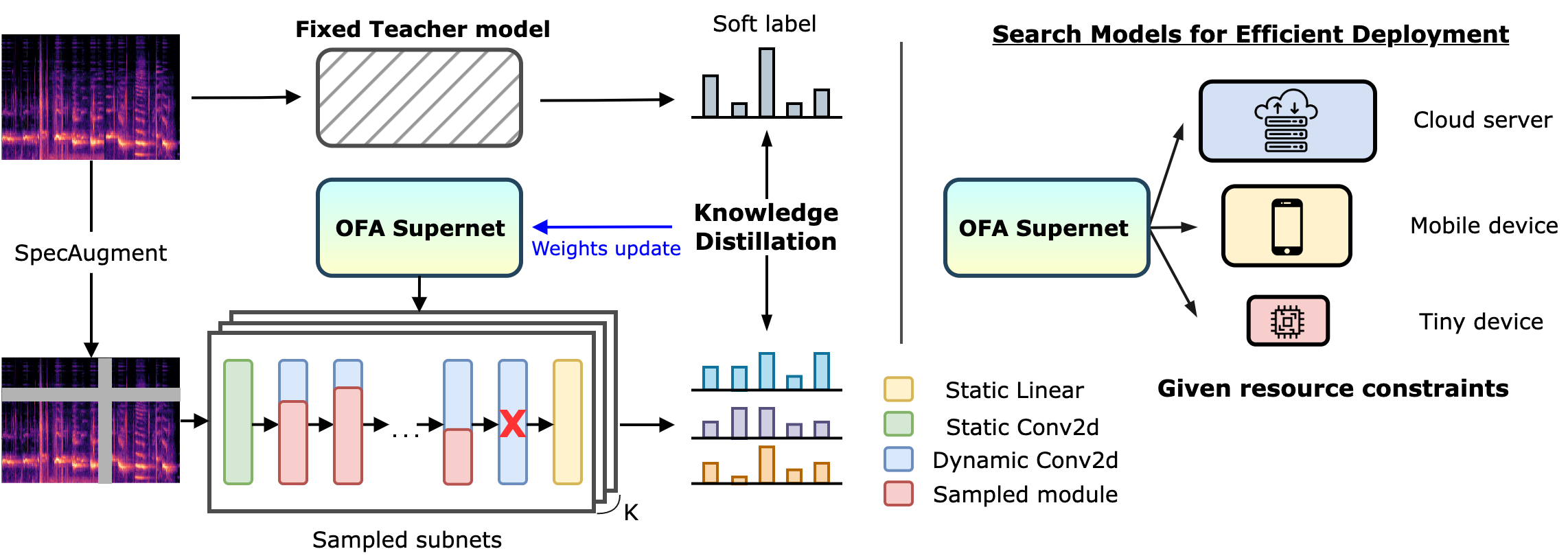}
  \vspace{-0.3cm}
  \caption{The overview of OFA AEC framework. \textit{Left}: During OFA training, we adopt a teacher model to generate soft labels. The students are $K$ sub-networks sampled from the supernet. We adopt knowledge distillation for training supernet, where only students' input spectrograms are distorted (we use SpecAugment). In the sampled subnets, the red-colored layers are with elastic-sampled channel width, and the cross sign means elastic discard layer. \textit{Right}: After supernet training, we search for the best-performing sub-network given specific device constraints. The searched models typically achieve good performance and do not require re-training.}
  \label{fig:overview}
\end{figure*}

\vspace{-0.3cm}
\section{Method}
This section introduces our Once-For-All for Acoustic Event Classification (OFA-AEC) framework. We would explain how we design neural architecture space, how we conduct training, and also the architecture search algorithm. 

\vspace{-0.3cm}
\subsection{Problem formulation}
In this work, we focus on acoustic event classification (AEC), a multi-label classification problem. 
We denote $x_n$ as an audio clip in our dataset where $n$ is the sample index, and $f(x_n) \in [0, 1]^C$ is the model output presenting the presence probability of the $C$ labels. $y_n \in \{0, 1\}^C$ represents the label of $x_n$. We would use knowledge distillation in our training, and we denote our trained teacher model (with superior performance) as $f_t$. 

As self-explained by its name, Once-for-All (OFA) training would train a supernet where sub-network architectures of this supernet would be used for diverse downstream deployments. We denote the weights of the OFA supernet as $w_o$, and $S(w_o, a)$ as the model selection scheme (via architectural configuration $a$) that selects the part of the supernet model with the configuration $a$. We assume there are in total $M$ supported configurations, and the selection scheme $S(w_o, a_i)$ would create the $i$-th supported sub-network.
The supernet training can be formulated as:
\vspace{-0.3cm}
\begin{equation}
   \mathop{min}_{w_o} \sum_{n=1}^{N} \sum_{i=1}^{M} \mathcal{L}(x_n, y_n; f_t, S(w_o, a_i)), 
   \label{eqn:ofa_1}
\end{equation}
where $\mathcal{L}$ is the knowledge distillation loss function.

\begin{figure}[t]
  \centering
\includegraphics[width=0.7\linewidth]{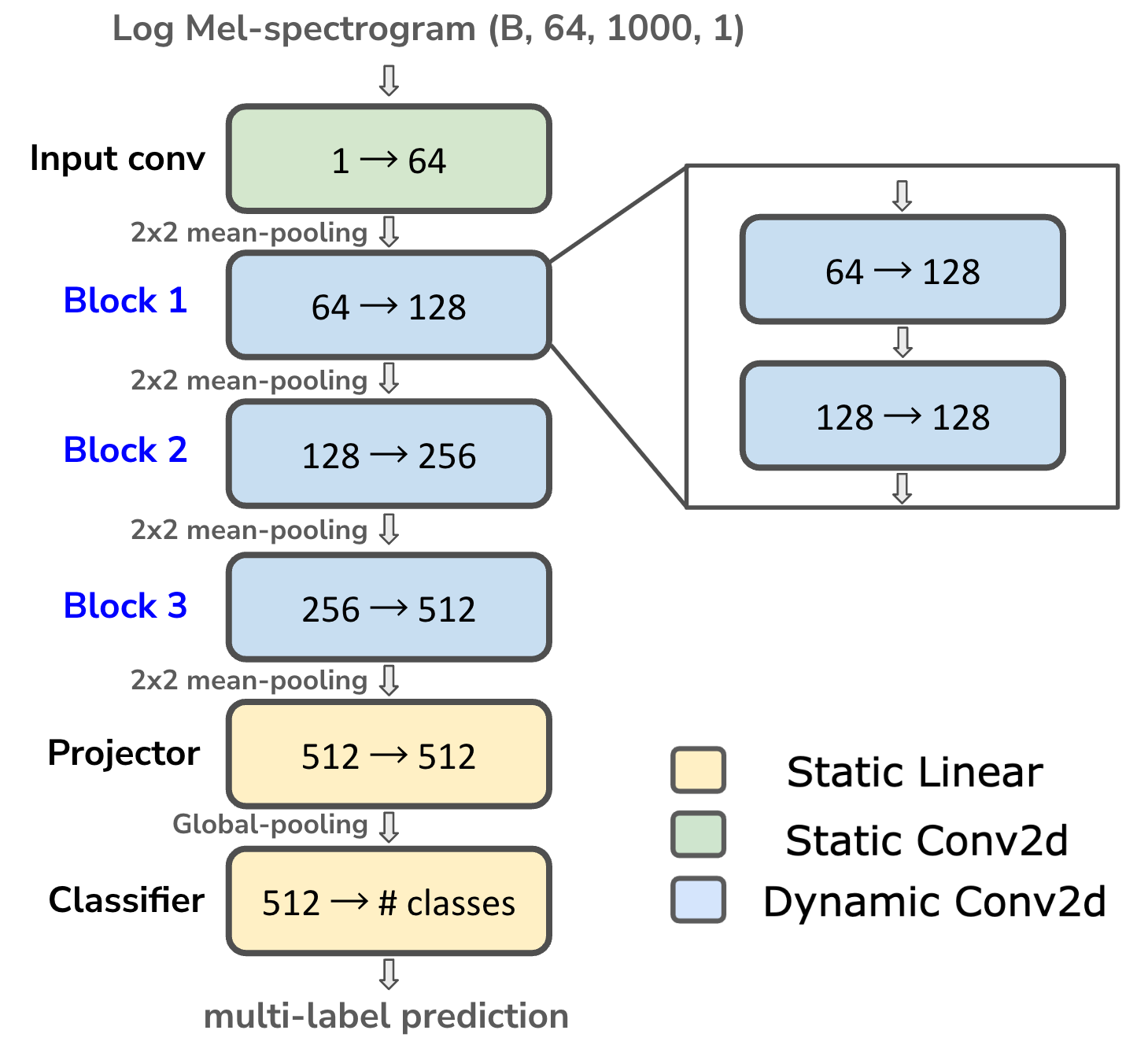}
  \caption{The architecture of the once-for-all supernet, which takes $64 \times 1000$ Mel spectrogram feature as input with batch size $B$. The value in blocks indicates the number of channels for CNN layers or feature size for feed-forward layers. 
  }
  \label{fig:model}
\end{figure}

\subsection{Architecture space}
\label{sec:2-2}
The architecture of the weight-sharing supernet is shown in Figure \ref{fig:model}. We divide a CNN-based AEC model into a sequence of blocks. The ``Static'' blocks refer to the standard fully-connected or convolutional 2D layers, whereas the ``Dynamic" block means it can be replaced by its sub-network. Each convolutional block contains 2 CNN layers, where an average-pooling layer with $2 \times 2$ pooling size is inserted between two consecutive blocks. On top of the final CNN layer, a global pooling layer, a fully-connected layer and a multi-label classifier with sigmoid activation functions are sequentially applied. 
All CNN layers use $3 \times 3$ kernel size.

The once-for-all supernet supports two searching dimensions for each block: 1) \textbf{Elastic Width (EW)}: \textit{arbitrary numbers of channels}. Given a width ratio $r$ and the full width size $D$, the layer uses the first $\floor{r \times D}$ channels. We experiment with four ratios, say $\{0.4, 0.6, 0.8, 1.0\}$. 2) \textbf{Elastic Depth (ED)}: \textit{arbitrary number of layers}. As one block has two layers, the dynamic depth for each block is either $1$ or $2$. 

We include both EW and ED when searching for sub-networks during supernet training; However, we observed that only applying ED in the last convolutional block is preferred according to the ablation study on search space design in Section \ref{ablation:space}. 

\subsection{Weight-sharing supernet training}
The basic idea for training a weight-sharing supernet is to uniformly sample a few sub-networks (e.g., $f_{s}^{i}$, $1\leq i \leq K$) from the $M$ supported sub-networks using the model selection scheme $S(w_o, a)$, then encourage the prediction of each sub-network to have high fidelity with respect to the teacher model $f_{t}$ via binary cross entropy loss.
The training objective (Equation \eqref{eqn:ofa_1}) can be approximated as
\begin{equation}
    \sum_{n=1}^{N} \sum_{i=1}^{K} BCE(f_{s}^{i}(x_n), f_t(x_n)),
\end{equation}
where $BCE(f_s(x_n), f_t(x_n)) =  f_t(x_n) \cdot \mathop{ln}f_s(x_n) + (1-f_t(x_n)) \cdot \mathop{ln}(1-f_s(x_n))$ is the knowledge distillation loss $\mathcal{L}$, and $K$ is the number of sampled neural networks and $N$ is the number of samples. 

\subsection{Search algorithm}
During inference, we adopt the random search to find the optimal sub-network under the specific resource constraint. Specifically, we search the best sub-networks given the number of parameters ranging from $[N_{params}-\epsilon, N_{params}+\epsilon]$, where $N_{params}$ is the specific constraint, and $\epsilon$ is the tolerance. We randomly sample $p$ sub-networks from the OFA supernet as the candidate sub-networks, run inference for those candidates on the validation set, and select the best one for each constraint.

\begin{table}[t]
\centering
\resizebox{0.48\textwidth}{!}{
\begin{tabular}{lcc|cc}
\toprule
\textbf{Model} & \textbf{\#Params(M)} & \textbf{\%} & \textbf{mAP} & \textbf{\%}   \\ 
\midrule
AST \cite{ast}            & 86.0  &   -  & 34.7 & - \\
SSAST \cite{ssast}          & 86.0  &  -   & 31.0 & -\\
\midrule
CNN10 (B = 32) \cite{pann}  & 5.2  & -   & 27.8 & -\\
CNN10 (B = 64)*  & 5.2  & 100.0   & \textbf{29.6} & 100.0 \\ 
\midrule
\midrule
Baseline & 1.0   & 19.2 & 21.2 & 71.6\\
KD      & 1.0   & 19.2 & 24.8 & 83.7\\
OFA        & 0.9   & 17.3 & \textbf{26.6} & \textbf{89.9}\\ 
\midrule
Baseline & 2.1   & 40.3 & 22.6 & 76.4\\
KD      & 2.1   & 40.3 & 26.4 & 89.2\\
OFA          & 1.8  &  34.6  & \textbf{27.2} & \textbf{91.9}\\ 
\midrule
Baseline & 3.5   & 67.3 & 24.8 & 83.8\\
KD     & 3.5   & 67.3 & \textbf{28.3} & \textbf{95.6}\\
OFA              & 3.6  & 69.2  & 27.9 & 94.3\\ 
\bottomrule
\end{tabular}
}
\vspace{-0.3cm}
\caption{Mean Average Precision (mAP) performance on AudioSet evaluation set. The training data is the balanced train set. ``B" denotes the batch size. ``*" is our \textit{teacher} model. ``\%" means the relative ratio compared to the teacher model.}
\label{tab:performance}
\end{table}

\vspace{-0.3cm}
\section{Experiments}
\subsection{Dataset}
We conduct our experiments on AudioSet \cite{audioset}, a well-known multi-label AEC benchmark corpus. There could be multiple events in a clip. The official balanced and evaluation sets are used as our train and test sets, with 22160 and 20371 10-second samples, respectively. The validation set is a subset randomly sampled from the unbalanced set with 28339 10-second samples without overlapping with the training set. The audio is pre-processed at 32K sampling rate.
We use log-mel spectrogram, $1000$ frames $\times$ 64 mel bins, as our input feature. There could be multiple events present in each 10-second clip, and thus this is a multi-label classification problem. We use well-adopted Mean Average Precision (mAP) as our evaluation metric.

\vspace{-0.2cm}
\subsection{The teacher model and Baslines}
The teacher model $f_t$ uses the CNN10 architecture designed in \cite{pann}. 
In order to achieve good performance, we train the teacher model using data augmentation techniques (e.g., mixup \cite{mixup} and specaugment \cite{specaugment}) up to 400K iterations using batch size of $64$ on AudioSet balanced train set. As shown in Table \ref{tab:performance}, our model achieves 29.6 mAP on the AudioSet evaluation set, which outperforms the CNN10 model (when also trained on balanced set) as shown in \cite{pann}. We compare OFA training with two baseline approaches: \\
\textbf{Baseline}: Training each model from scratch using BCE loss and specaugment, which is a standard way to train an AEC model.\\
\textbf{Knowledge Distillation (KD)}: Training each model with knowledge distillation and specaugment, which is a stronger baseline. 

For the architecture of individual models, based on the supernet search space, we set each model with a fixed width ratio and depth across all the dynamic blocks. For example, ``width ratio $= 0.4$, depth $= 2$" for all three dynamic CNN blocks. Thus, the individual models are all with a double-increasing number of channels. Note that the architectures of these individually-trained models are also supported architecture configurations in our search space. 

\vspace{-0.4cm}
\subsection{Implementation details}

We use the batch size of 64, and Adam optimizer with a 0.001 learning rate for all the experiments. Our specaugment \cite{specaugment} consists of two time and frequency masks on the spectrogram ($T=64, F=8$). For supernet training, we first only train the largest sub-network for 100K iterations, then train the sampled sub-networks for 200K iterations. The number of sampled sub-networks $K$ is 4. For the baseline methods, we train each model with 100K iterations. As the number of sub-network parameters ranges from 0.6M to 5.2M, we choose four computational constraints $\{0.8M, 1.8M, 2.8M, 3.8M\}$ with $\epsilon=0.2M$ tolerance for the random search. The population size $p$ for random search is 25.

\vspace{-0.4cm}
\subsection{Results}
The main results are shown in Table \ref{tab:performance}. First, for the models of size at ~1M \#Params, our OFA method yields significantly better performance than the baselines. Our searched model retains 89.9\% performance of the teacher model with only 17.3\% \#Params. On the other hand, training from scratch and knowledge distillation models achieve 21.2 and 24.8 mAP, respectively. 
The OFA method also achieves superior performance at around 2M \#Params, surpassing the baseline by 4.6 mAP and the KD method by 0.8 mAP. The results demonstrate that the weight-sharing NAS has the capability to produce small networks with outstanding performance. 
For the larger model size at around 3.5M \#Params, the OFA method achieves 27.9 mAP, which is much better than the baseline method (24.8 mAP) but slightly worse than the knowledge distillation model (28.3 mAP). 
Overall, by just training one weight-sharing supernet, we can search for sub-networks that have superior performance than baselines. Even when compared with KD, our methods also achieve at least comparable and usually superior performance. Please note, it is possible to achieve better performance if we use an advanced search algorithm instead of random search.

\begin{figure}[t]
  \centering
\includegraphics[width=0.8\linewidth]{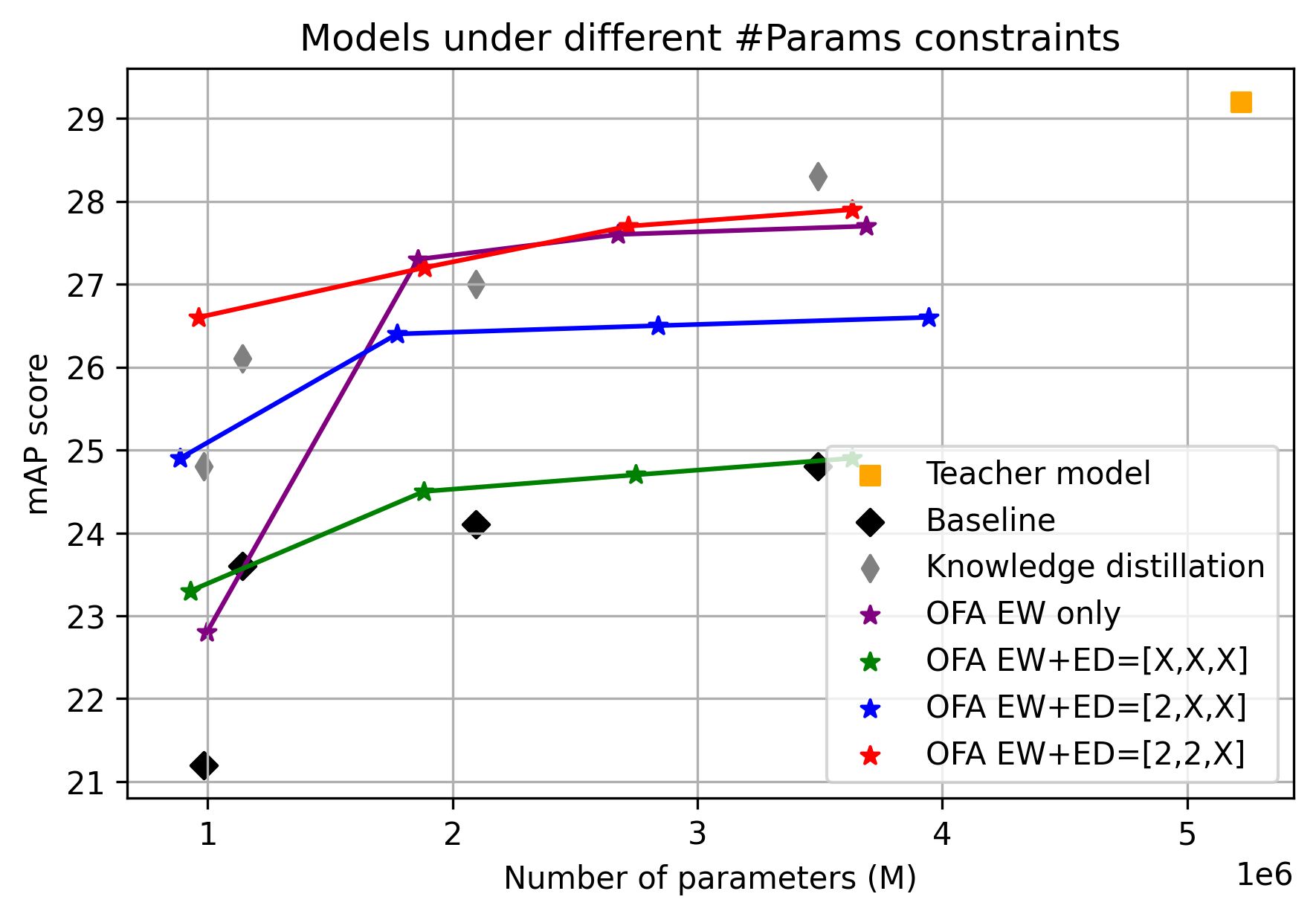}
  \vspace{-0.4cm}
  \caption{Comparison of different search spaces for the OFA supernet. We selectively apply ED on certain blocks, indicated with ``X". }
  \label{fig:ablation_ED}
\end{figure}
\section{Discussion}
\subsection{Ablation study of search space}
\label{ablation:space}
We try different search spaces for the OFA-AEC supernet, showing the performance of the searched sub-networks in Figure \ref{fig:ablation_ED}. We always use EW, and selectively use ED on certain blocks (represented as ``X" in Figure \ref{fig:ablation_ED}). We first experiment with using ED for \textit{all} the dynamic blocks (Figure \ref{fig:ablation_ED} ``OFA EW+ED=[X,X,X]" curve) and it shows significant degradation in performance. To explore suitable ways for using ED, we try not applying elastic depth in some blocks (denoted as ``2" since all the two layers in that block would be retained). We find out that applying elastic depth to the front blocks is detrimental while applying ED to the later blocks is preferred. Especially, applying ED to the last convolutional block (``OFA EW+ED=[2,2,X]" curve) achieves superior performance, especially for ultra-small models (e.g., 26.6 mAP with 0.9M \#Params). 

We also investigate only applying elastic width, denoted as \textit{EW-only}. In addition to the four width ratios, we add the minimal width ratio of 0.2 to enable the EW-only supernet containing small models around 1M \#Params. EW-only supernet achieves comparable performance as the EW+ED=[2,2,X] when the model size exceeds 2M. However, it performs poorly at the scale of 1M \#Params (22.8 mAP). These results indicate the necessity of using both EW and ED in supernet training and search, but we need to be careful when applying ED.

The sensitivity of elastic depth has not been discussed and observed in previous OFA methods on image recognition. Compared to their OFA model, which is a deep CNN model with several residual connection blocks, our AEC model is much shallower and based on standard convolutional layers. Hopefully, our ablation study can provide more insights to apply OFA training on architectures with different levels of complexity and on other audio tasks.

\begin{figure}[t]
  \centering
\includegraphics[width=0.8\linewidth]{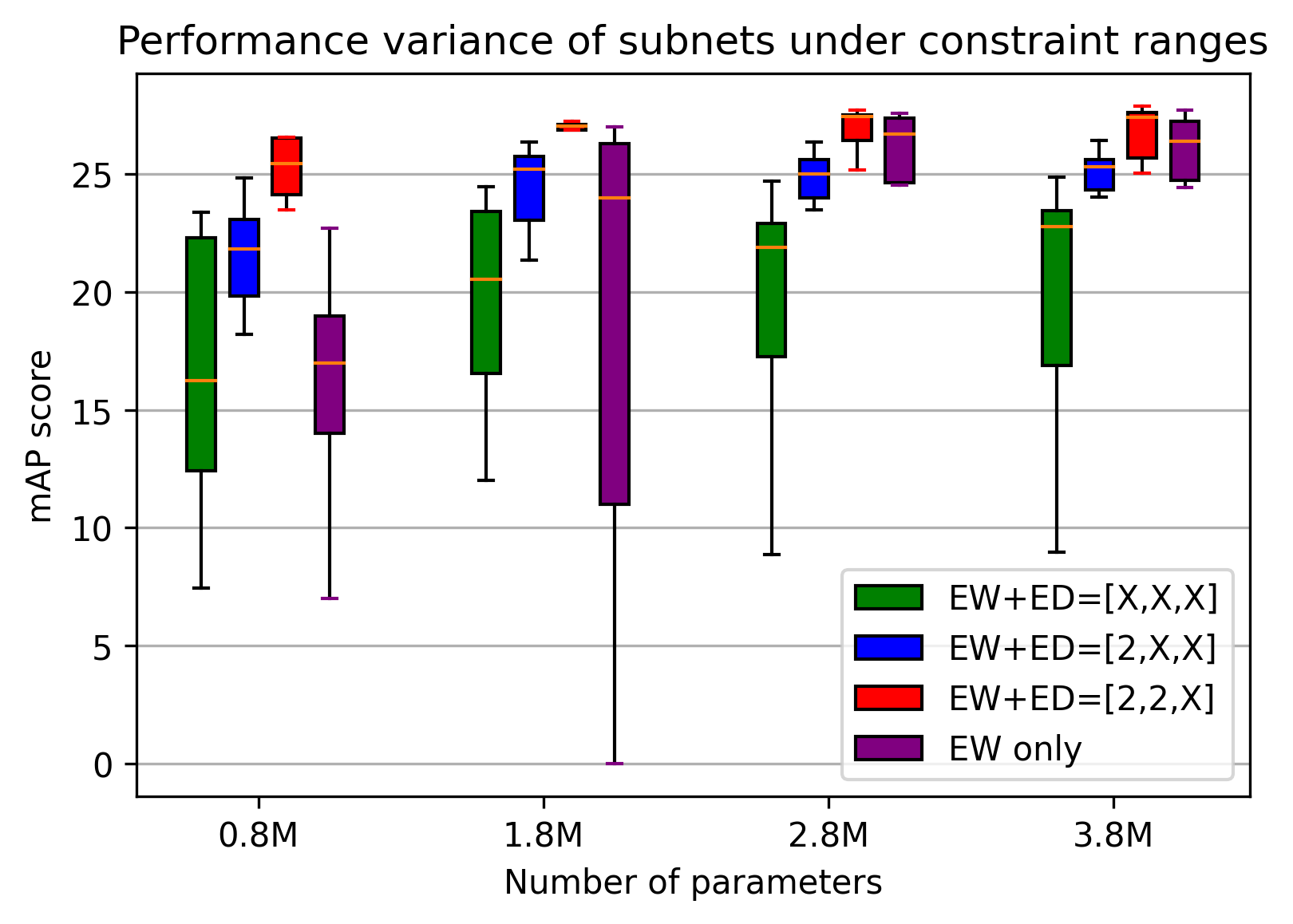}
  \vspace{-0.4cm}
  \caption{The boxplot of candidate sub-network performance with four \#Params constraints. The mean value is shown in the orange line. We selectively apply ED on certain blocks, indicated with ``X".}
  \label{fig:variance}
\end{figure}

\vspace{-0.3cm}
\begin{table}[t]
\centering
\begin{tabular}{cccc|c}
\toprule
\#Params(M) & OFA   & +10K & +100K & Init+100K\\ \midrule
0.9         & 26.6 &  26.7  & 26.7 & 24.8 \\
1.8         & 27.2 &  27.6  & 27.8 & 26.6 \\
2.7         & 27.7 &  28.2  & 28.4 & 28.1 \\
3.6         & 27.9 &  28.3  & 28.5 & 28.3 \\ \bottomrule
\end{tabular}
\vspace{-0.2cm}
\caption{The results of continual fine-tuning for 10K or 100K iterations on the OFA searched sub-networks. The OFA method here is EW+ED=[2,2,X]. ``Init" means we used the identical sub-network architecture found by OFA, and trained with knowledge distillation.}
\label{tab:CF}
\end{table}

\subsection{Performance variance of sub-networks}
We measure the performance variance of sub-networks with similar model sizes. We include different search spaces of supernets for analysis. The result is shown in Figure \ref{fig:variance}. We observe that if using elastic depth in all blocks (green boxes), the average performance is the worst, and the standard deviation is the largest across four \#Params constraints. The large variance implies the instability of supernet training and searching. In contrast, only using elastic depth in the last block (red boxes) achieves the best average performance with low variance, showing that the supernet converges well and is robust for searching. Lastly, though EW-only supernet achieves good performance (see Figure ~\ref{fig:ablation_ED}), the performance variance among different sub-network is large, especially at 1.8M \#Params constraints. 

\subsection{Continual fine-tuning}
To evaluate whether the searched sub-networks are well-trained, we continually fine-tune (with knowledge distillation) the searched sub-networks for 10K and 100K iterations. To better understand OFA from the perspective of optimization, we also train the same architectures of the searched sub-networks from scratch using knowledge distillation and show the results in column ``Init+100K".

According to the results shown in Table \ref{tab:CF}, continually fine-tuning the OFA-identified sub-networks generally helps improve the performance except for the models at 0.9M \#Params (we only observed 0.1 mAP improvement after fine-tuning). In the meantime, the OFA-trained model significantly outperforms the KD-trained counterpart at 0.9M \#Params. These observations suggest that the ultra-small sub-networks are already well-trained. When looking at models at 3.6M \#Params, the OFA-trained sub-networks only achieve on-par performance compared with KD-trained individual models, which aligns with Table ~\ref{tab:performance}. However, continual training does improve the sub-networks at 3.6M \#Params by 0.6 mAP, and finally outperform the KD counterpart.

This analysis shows that the weight-sharing supernet training strategy provides a significant advantage in terms of optimization over training a single model either from scratch or via KD. For ultra-small models, OFA-trained sub-networks can already significantly outperform KD-trained counterparts; For larger model size, OFA-trained sub-networks weights can serve as good initialization and achieves superior (compared with KD) performance after continual fine-tuning. Thus, the main benefit of OFA comes not only from searching for better architecture but also from optimization.

\vspace{-0.3cm}
\section{Conclusion}
Efficiently training models satisfying diverse on-device computation resource constraints becomes an emerging topic in the industry as edge computing is gaining popularity. In this work, we apply the once-for-all framework to address this issue for acoustic event classification. To the best of our knowledge, we propose the first weight-sharing NAS method for CNN-based AEC models. 
In addition to the saving in model training efforts brought by the once-for-all training, the performance of searched sub-networks across different computation resource constraints is superior to models trained from scratch and models trained with knowledge distillation. 
Moreover, we analyze the search space design for the OFA AEC model and find that the benefit of weight-sharing supernet training of ultra-small models does not only come from searching, but to a large extent, comes from optimization.
In the future, we plan to explore other AEC model architectures for supernet and extend our framework to other audio applications.


\bibliographystyle{IEEE}
\bibliography{refs}

\end{document}